# Effect of Magnetic Field on Neutral Bath Containing Charged Brownian Particles

Jana Tóthová[1], Ján Buša[2], and Vladimír Lisý[1]

[1] Faculty of Electrical Engineering and Informatics, Technical University of Košice, 04200 Košice, Slovakia
[2] Meshcheryakov Laboratory of Information Technologies, Joint Institute for Nuclear Research, Dubna, 141980 Russia

**Based on the Zwanzig-Caldeira-Legget theory generalized to systems under the influence of a static magnetic field, we obtain equations of motion for the Brownian particle (BP) and oscillators constituting the bath in which the BP is embedded. The equations are of the type of a generalized Langevin equation, which accounts for the frictional memory of the system. The BP is assumed to be charged while the bath particles are neutral. They thus do not directly respond to the external field, but their interaction with the BP leads to changes in the bath state. Using the solution of the equations found, we calculate the average bath angular momentum and show that it persists for long times when the system is assumed to reach equilibrium. This indicates a possible violation of the Bohr-van Leeuwen theorem for baths consisting of charged particles. However, this must be confirmed by a substantial generalization of the presented model when the bath particles feel the external field, which affects the memory in the dynamics of the system.**

*Index Terms*—Angular momentum, charged Brownian particle, external magnetic field, neutral bath.

## I. Introduction

In the last two decades, the Brownian motion (BM) of charged particles in external magnetic fields has attracted considerable attention, mainly due to the necessity of taking into account memory effects in the particle dynamics [1]-[7]. It has been shown that the consideration of the correlated (colored) thermal noise leads to several peculiarities not possessed by the theories in which the Brownian particles (BPs) are driven by the white noise. For example, the mean square displacement of the BP, in addition to the 'Einstein' term proportional to the time, contains other nontrivial contributions. Here, we develop these works and consider a problem that is closely related to the famous Bohr-van Leeuwen (BvL) theorem [8], [9], which governs the response of equilibrium systems to external magnetic fields. This theorem, stating the absence of classical magnetization, is expected to be valid also for systems of charged BPs immersed in a bath of other particles. In a comprehensive work by Matevosyan and Allahverdyan [10], such a problem was already studied assuming that the bath is constituted of neutral particles. The used model and the equations of motion for the BP and the bath are thus a special case of the theory [6, 7], where both the BP and the bath particles were charged. It has been found in [10] that in a static magnetic field, the bath particles, due to their interaction with the BP, acquire a nonzero angular momentum, which is a very intriguing result. We suppose, however, that this work should be revisited, mainly because it assumes that the magnetic moment of the BP induced by the static external magnetic field is zero. As opposite, we have shown, by analytical calculations and precise numerical solution of the model, that this is not true: the angular momentum of a charged classical BP in a neutral bath is nonzero even at long times of observation when the system should be in equilibrium. At infinite times, the bath exhibits a diamagnetic moment described by a simple formula [11]. Here, we use this formula and calculate the angular momentum of the bath particles. Although the particles are neutral, the long-time momentum is nonzero. This coincides with the conclusion of the work [10]; however, our final result significantly differs, although we use the same Zwanzig-Caldeira-Legget (ZCL) model [12]-[14], which is frequently and effectively used in the theory of the BM.

## II. The Zwanzig-Caldeira-Legget Model for the BM in a Static Magnetic Field

We come from the ZCL Hamiltonian describing the equilibrium state up to the moment $t = 0$ when the external magnetic field $\mathbf{B} = (0,0,B)$ is switched on,

$$H = \frac{\mathbf{p}^2}{2m} + \frac{1}{2}\sum_{i=1}^{N}\left[\frac{\mathbf{p}_i^2}{m_i} + m_i\omega_i^2\left(\mathbf{r}_i - \frac{c_i\mathbf{r}}{\omega_i^2}\right)^2\right]. \quad (1)$$

Here $\omega_i$ are the eigenfrequencies of the bath particles (which are oscillators) of masses $m_i$, positions $\mathbf{r}_i(t)$, and momenta $\mathbf{p}_i(t)$. They surround the BP of mass $m$, position $\mathbf{r}(t)$, and momentum $\mathbf{p}(t)$. The constants $c_i$ characterize the strength of their interaction with the BP. For identical particles, we will use $m_i \equiv \mu$ and $c_i \equiv c$. By a standard way, using the equations $\dot{x}_i = p_{xi}/m_i$ and $\dot{p}_{xi} = -\partial H/\partial x_i$, and similarly for other projections, one obtains equations for the bath particles and the BP. After the inclusion of the magnetic force, they have the following form:

$$\dot{\mathbf{r}} = \frac{\mathbf{p}}{m}, \quad \dot{\mathbf{p}} = \sum_i m_i c_i\left(\mathbf{r}_i - \frac{c_i}{\omega_i^2}\mathbf{r}\right) + Q\dot{\mathbf{r}}\times\mathbf{B}, \quad (2)$$

Corresponding author:
J. Tóthová (e-mail: jana.tothova@tuke.sk).

$$\dot{\mathbf{r}}_i = \frac{\mathbf{p}_i}{m_i}, \qquad \dot{\mathbf{p}}_i = -m_i \omega_i^2 \mathbf{r}_i + m_i c_i \mathbf{r}. \qquad (3)$$

Our interest is in the motion of the BP in the plane $x$, $y$, since its motion along the axis $z$ is not affected by the magnetic field. From (3), using the Laplace transformation (LT) $\mathcal{L}\{\varphi(t)\} = \tilde{\varphi}(s) = \int_0^\infty \varphi(t) e^{-st} dt$ (the inverse LT will be denoted as $\mathcal{L}^{-1}\{\tilde{\varphi}(s)\} = \varphi(t)$), we obtain the algebraic equation for the $x_i$ projection of $\mathbf{r}_i$,

$$s^2 \tilde{x}_i(s) - sx_i(0) - \dot{x}_i(0) = -\omega_i^2 \tilde{x}_i(s) + c_i \tilde{x}(s). \qquad (4)$$

Its solution, in the time domain, has the form

$$x_i(t) = x_i(0)\cos(\omega_i t) + \frac{\dot{x}_i(0)}{\omega_i}\sin(\omega_i t) \\ + \frac{c_i}{\omega_i}\int_0^t x(t')\sin[\omega_i(t-t')]dt'. \qquad (5)$$

In (4) and (5), $x_i(0)$ and $\dot{x}_i(0)$ are the positions and velocities of the oscillators at $t = 0$, and $x(0)$ is the initial position of the BP. The quantities $x_i(0)$, $\dot{x}_i(0)$, and $x(0)$ are random with zero means. The same equations hold for $y$ projections. After substitution of the solution (5) for $x_i$ (and the same for $y_i$) in (2), one gets the generalized Langevin equation (GLE) for the $x$-projection of the velocity of the BP

$$m\dot{\upsilon}_x(t) = QB\upsilon_y(t) - \int_0^t \upsilon_x(t')\Gamma(t-t')dt' + f_x(t). \qquad (6)$$

The equation for the $y$-projection differs only by the change of $x$ to $y$ and the sign of the first term on the right-hand side. With the help of the convolution theorem and using the rule for the LT of a differentiated function [15], the solutions of these equations are

$$\tilde{x}(s) = \frac{x(0)}{s} + m\dot{x}(0)\tilde{\Phi}(s) + m\dot{y}(0)\tilde{\Psi}(s) + \tilde{\Phi}(s)\tilde{f}_x(s) + \tilde{\Psi}(s)\tilde{f}_y(s),$$

$$\tilde{y}(s) = \frac{y(0)}{s} + m\dot{y}(0)\tilde{\Phi}(s) - m\dot{x}(0)\tilde{\Psi}(s) + \tilde{\Phi}(s)\tilde{f}_y(s) - \tilde{\Psi}(s)\tilde{f}_x(s),$$
$$(7)$$
$$\tilde{\upsilon}_x(s) = m\dot{x}(0)s\tilde{\Phi}(s) + m\dot{y}(0)s\tilde{\Psi}(s) + s\tilde{\Phi}(s)\tilde{f}_x(s) + s\tilde{\Psi}(s)\tilde{f}_y(s),$$

$$\tilde{\upsilon}_y(s) = m\dot{y}(0)s\tilde{\Phi}(s) - m\dot{x}(0)s\tilde{\Psi}(s) + s\tilde{\Phi}(s)\tilde{f}_y(s) - s\tilde{\Psi}(s)\tilde{f}_x(s),$$
$$(8)$$

where

$$\tilde{\Phi}(s) = \frac{ms^2 + s\tilde{\Gamma}(s)}{[ms^2 + s\tilde{\Gamma}(s)]^2 + (QBs)^2}, \qquad (9)$$

$$\tilde{\Psi}(s) = \frac{QBs}{[ms^2 + s\tilde{\Gamma}(s)]^2 + (QBs)^2}. \qquad (10)$$

The integral in (6) generalizes the Stokes friction force proportional to the velocity, and the kernel $\Gamma(t) = \sum_i m_i c_i^2 \omega_i^{-2} \cos(\omega_i t)$ is the so-called memory function. It is connected to the random force

$$f_x(t) = \sum_{i=1}^N m_i c_i \left\{ [x_i(0) - \frac{c_i}{\omega_i^2}x(0)]\cos(\omega_i t) + \frac{\dot{x}_i(0)}{\omega_i}\sin(\omega_i t)\right\},$$
$$(11)$$

through the fluctuation-dissipation theorem [16] by the relation $\langle f_x(t)f_y(t')\rangle = k_B T \delta_{xy}\Gamma(|t-t'|)$, if the system is stationary.

Depending on the system parameters and the distribution of the frequencies $\omega_i$, the memory functions $\Gamma(t)$ can be very different. In the simplest case of white thermal noise $f(t)$, as in the original Langevin theory of the BM [17], we have $\Gamma(t) = 2\xi\delta(t)$, where $\xi$ is the friction coefficient. A popular generalization of the classical Langevin equation assumes an exponentially decaying memory in time (corresponding to the so-called Ornstein-Uhlenbeck noise), $\Gamma(t) = (\xi/\tau)\exp(-t/\tau)$, with the LT $\tilde{\Gamma}(s) = \xi/(\tau s + 1)$. One returns to the Langevin theory when the relaxation time $\tau$ converges to zero. The exponentially relaxing memory function can be within the ZCL model obtained assuming that the number of particles is very large and the frequencies can change from 0 to ∞, so that it is possible to transform the sums to integrals, $\sum_i F(\omega_i) \to \int F(\omega)h(\omega)d\omega$ [13], with the distribution of the Drude type, $h(\omega) = (2\xi\omega^2/\pi\mu\tau^2 c^2)/(\omega^2 + \tau^{-2})$.

### III. ANGULAR MOMENTUM OF THE BATH PARTICLES

We aim to calculate the mean angular momentum of the bath particles, $m_i\langle L_i(t)\rangle$ at $t \to \infty$, determined by $L_i(t) = x_i(t)\upsilon_{iy}(t) - y_i(t)\upsilon_{ix}(t)$. We will need the correlators of the initial positions and velocities. In the ZCL model, the bath oscillators do not interact so that the correlators of different particles are zero. Also, the $xy$ cross-correlators of the initial positions and velocities, and the correlators between the bath particles and the initial position of the BP can be skipped. The nonzero mean values follow from the equilibrium Hamiltonian (1) with the use of the distribution $h(\omega)$:

$$\langle x_j^2(0)\rangle = k_B T / m_j \omega_j^2, \quad \langle \dot{x}_j^2(0)\rangle = k_B T / m_j,$$
$$(12)$$
$$\langle x^2(0)\rangle = k_B T \tau / \xi, \quad \langle \upsilon_x^2(0)\rangle = k_B T / m$$

(the same means hold after replacing $x$ to $y$ and for $\langle x^2(0)\rangle$ we already used that the bath particles are identical).

If one skips in $L_i(t)$ the terms that do not contribute to its average value, the remaining quantities to be considered in

$x_i(t)v_{iy}(t)$ are, from (5) and the accompanying equations for $y$ projections, as follows.

$$I = -c_i \sin(\omega_i t) y_i(0) \int_0^t x(t') \sin[\omega_i(t-t')]dt',$$

$$II = \frac{c_i}{\omega_i} \cos(\omega_i t) \dot{y}(0) \int_0^t x(t') \sin[\omega_i(t-t')]dt',$$

$$III = c_i \cos(\omega_i t) \int_0^t y(t') \cos[\omega_i(t-t')]dt',$$

$$IV = \frac{c_i}{\omega_i} \sin(\omega_i t) \dot{x}(0) \int_0^t x(t') \cos[\omega_i(t-t')]dt',$$

$$V = \frac{c_i^2}{\omega_i} \int_0^t x(t') \sin[\omega_i(t-t')]dt' \int_0^t y(t'') \cos[\omega_i(t-t'')]dt''.$$

The terms forming $y_i(t)v_{ix}(t)$ are the same if all $x$ are replaced by $y$. Let us now calculate the mean values of $I - V$.

In $I$, only the term with $f_y(t')$ in $x(t')$ that in the sum contains $y_i(0)$ will contribute to $\langle I \rangle$, giving after the averaging $m_i c_i \cos(\omega_i t') \langle y_i^2(0) \rangle$ with $\langle y_i^2(0) \rangle$ from (12). An analogous term to $\langle I \rangle$, but with the sign "–" follows from (5) and (7) with $y \to x$ and vice versa, taking into account that $\langle x_i^2(0) \rangle = \langle y_i^2(0) \rangle$. The sum of the two terms gives

$$\langle I \rangle = \frac{-2k_B T c_i^2}{\omega_i^2} \sin(\omega_i t) \int_0^t dt' \sin[\omega_i(t-t')] \mathcal{L}^{-1}(t') \left\{ \frac{s\tilde{\Psi}(s)}{s^2+\omega_i^2} \right\},$$

where we again used the convolution theorem for the LT.

In the same way, we find the terms $\langle II \rangle - \langle IV \rangle$ determined by the means (12):

$$\langle II \rangle = \frac{2k_B T c_i^2}{\omega_i^2} \cos(\omega_i t) \int_0^t dt' \sin[\omega_i(t-t')] \mathcal{L}^{-1}(t') \left\{ \frac{\tilde{\Psi}(s)}{s^2+\omega_i^2} \right\},$$

$$\langle III \rangle = -\frac{2k_B T c_i^2}{\omega_i^2} \cos(\omega_i t) \int_0^t dt' \cos[\omega_i(t-t')] \mathcal{L}^{-1}(t') \left\{ \frac{s\tilde{\Psi}(s)}{s^2+\omega_i^2} \right\},$$

$$\langle IV \rangle = -\frac{2k_B T c_i^2}{\omega_i^2} \sin(\omega_i t) \int_0^t dt' \cos[\omega_i(t-t')] \mathcal{L}^{-1}(t') \left\{ \frac{\tilde{\Psi}(s)}{s^2+\omega_i^2} \right\}.$$

The behavior of $\langle I \rangle - \langle IV \rangle$ can be most easily determined by transforming the integrals in the LT and using the series expansions in small $s$ that correspond to long times. Only the integral in $\langle II \rangle$ remains nonzero as $t \to \infty$ and we obtain

$$\langle II \rangle = 2k_B T \frac{c_i^2}{\omega_i^4} \frac{QB}{\xi^2 + (QB)^2} \cos(\omega_i t). \quad (13)$$

Now we turn to the most involved calculation of the term $\langle V \rangle$. To sum the parts corresponding $x_i(t)v_{iy}(t)$ and $y_i(t)v_{ix}(t)$, one can use the identity

$\langle x(t')y(t'') \rangle = -\langle x(t'')y(t') \rangle$ [10]. In the LT, $V$ can be expressed as

$$V = 2\frac{c_i^2}{\omega_i} \mathcal{L}_t^{-1} \left\{ \frac{\omega_i}{\omega_i^2 + s^2} \tilde{x}(s) \right\} \mathcal{L}_t^{-1} \left\{ \frac{s}{\omega_i^2 + s^2} \tilde{y}(s) \right\}.$$

Having in mind the averaging of this expression over the initial values of the positions and velocities of the BP and with the random forces, we keep only the nonzero terms containing the means (12), and, in the time domain, include the correlator of the forces $\langle f_x(t) f_x(t') \rangle = (k_B T \tau / \xi) \exp(-|t-t'|/\tau)$. The straightforward but rather cumbersome calculations, including the expansions in $s \to 0$ as above for $\langle I \rangle - \langle IV \rangle$, give zero values for terms containing $\dot{x}(0)$ and $\dot{y}(0)$. One of the nonzero terms is $\sim \langle x^2(0) \rangle = \langle y^2(0) \rangle$ that contains, if the sum is replaced by an integral, $s \sum_i m_i c_i^2 / [\omega_i^2(\omega_i^2 + s^2)]$ $\approx \mu c^2 s \int_0^\infty d\omega h(\omega) / [\omega^2(\omega^2 + s^2)] = \xi/(1+\tau s)$. The result for this term, which we denote as $\langle V_1 \rangle$, is $\langle V_1 \rangle \approx -2k_B T c^2 \omega^{-3}$ $\cdot QB\tau \sin(\omega t) [\xi^2 + (QB)^2]^{-1} \mathcal{L}_t^{-1}\{1/s + \text{const} + O(s)\}$, so that at long times we have

$$\langle V_1 \rangle \approx -2k_B T \frac{c^2}{\omega^3} \frac{QB\tau}{\xi^2 + (QB)^2} \sin(\omega t). \quad (14)$$

The last contribution to $\langle V \rangle$ is determined by the terms each containing the random force,

$$\frac{V_2}{2c_i^2} = \mathcal{L}_t^{-1} \left\{ \frac{s\tilde{\Phi}(s)}{\omega_i^2 + s^2} \tilde{f}_y(s) \right\} \mathcal{L}_t^{-1} \left\{ \frac{\tilde{\Psi}(s)}{\omega_i^2 + s^2} \tilde{f}_y(s) \right\}$$
$$- \mathcal{L}_t^{-1} \left\{ \frac{\tilde{\Phi}(s)}{\omega_i^2 + s^2} \tilde{f}_x(s) \right\} \mathcal{L}_t^{-1} \left\{ \frac{s\tilde{\Psi}(s)}{\omega_i^2 + s^2} \tilde{f}_x(s) \right\}.$$

It was taken into account that the different projections of the random force do not correlate. Using again the expansions in small $s$, the limit of $\langle V_2 \rangle$ at $t \to \infty$ is given by the expression

$$\lim_{t \to \infty} \langle V_2(t) \rangle = \frac{2k_B T c^2 \xi}{\omega^4 \tau} \int_0^t dt' \int_0^t dt'' [\Psi(t')\Phi'(t'')$$
$$- \Psi'(t')\Phi(t'')] \exp\frac{-|t'-t''|}{\tau} \quad (15)$$

The inverse LT was calculated using that the initial values of the functions $\Phi(t)$ and $\Psi(t)$ are zero, so that $\mathcal{L}\{\Phi'(t)\} = s\tilde{\Phi}(s)$ and $\mathcal{L}\{\Psi'(t)\} = s\tilde{\Psi}(s)$. Equation (15) is exactly (except the factor $c^2/\omega^4$) the same as the one obtained in [11] for the quantity $\lim_{t \to \infty} \langle L(t) \rangle$ that determines the full angular momentum of a charged BP in a static magnetic field. With this result, (15) reads

$$\lim_{t\to\infty}\langle V_2(t)\rangle = -2\frac{k_B T c^2}{\omega^4}\frac{QB}{(QB)^2+\xi^2}. \quad (16)$$

Putting together the partial results (13), (14), and (16), the final formula for the mean long-time angular momentum $\mu\langle L(t)\rangle$ of a bath oscillator with frequency $\omega$ is given by

$$\frac{\langle L(t)\rangle}{D(c\tau^2)^2} = \frac{1}{(\omega\tau)^4}\frac{QB\xi}{\xi^2+(QB)^2}\left[\cos(\omega t)-\omega\tau\sin(\omega t)-1\right], \quad (17)$$

where $D = 2k_B T/\xi$ is the Stokes–Einstein–Sutherland diffusion coefficient for the BP in the plane $x$, $y$. Note that $\langle L(t)\rangle$ changes the sign as $\omega t$ grows (it is positive for $\omega\tau < \cot\omega t - \csc\omega t$). There is no time-independent limit of this quantity: $\langle L(t)\rangle$ has an oscillatory character even for very long times. Note also the following significant moment in the derivation of (17). As distinct from a number of papers in which the initial values of the positions and velocities of the particles in the system are set to zero (see, e.g., the work [18] on the theme), we consider these values nonzero. Within the ZCL model, these random values determine the thermal noise and the memory function. If they were neglected, the final result significantly differs from the proposed result (17).

Finally, with the use of the frequency distribution function $h(\omega)$, one can get the angular momentum of the whole bath summing all the contributions (17) of individual bath oscillators. This gives a simple formula

$$\mu\langle L(t)\rangle_{\text{bath}} = \mu\int_0^\infty \langle L(t)\rangle h(\omega)d\omega = -\frac{QB}{1+(QB/\xi)^2}Dt. \quad (18)$$

## IV. Conclusion

In conclusion, we have shown that the ZCL model used to describe the dynamics of a bath in which a charged BP is immersed leads to a surprising result of nonzero angular momentum of neutral bath particles (oscillators) when the system is put in a static magnetic field. The angular momentum persists for long times after the external field is switched on, when the system is expected to reach equilibrium. For an individual bath particle, it has an oscillatory character and disappears at very strong fields, which is another counterintuitive result. The summation of the angular momenta of all bath oscillators gives an exceedingly simple formula for $\mu\langle L(t)\rangle$: it does not depend on the particle mass $\mu$ and is proportional to the Einstein mean square displacement of the particle in the plane perpendicular to the magnetic field. Thus, although the bath particles do not directly respond to the magnetic field, their interaction with the BP, whose dynamics is affected by the field, leads to a change of the bath state. One could speculate whether this effect indicates a possible violation of the Bohr-van Leeuwen theorem for baths consisting of charged particles that states the absence of a magnetic moment induced by a static magnetic field. However, this must be confirmed by a substantial generalization of the presented model when the bath particles feel the external field, which affects memory in the dynamics of the system.

## Acknowledgment

This work was supported by the Scientific Agency of the Slovak Republic through grant VEGA 1/0353/22.